\documentstyle[12pt]{article}
\begin{document}
\thispagestyle{empty}

\begin{center}
               RUSSIAN GRAVITATIONAL ASSOCIATION\\
               CENTER FOR SURFACE AND VACUUM RESEARCH\\
               DEPARTMENT OF FUNDAMENTAL INTERACTIONS AND METROLOGY\\
\end{center}
\vskip 4ex
\begin{flushright}                 RGA-CSVR-012/94\\
                                   gr-qc/9408004
\end{flushright}
\vskip 15mm

\begin{center}
     {\large\bf
Dynamics of inhomogeneities of metric \\
\vskip1ex
in the vicinity of a singularity \\
\vskip1ex
in multidimensional cosmology }\\

\vskip2.5ex
     {\bf A.A.Kirillov, V.N.Melnikov$^{\dag}$}\\

\vskip 5mm
     $^{\dag}${\em Center for Surface and Vacuum Research,\\
     8 Kravchenko str., Moscow 117331, Russia}\\
     {\em Institute for Applied Mathematics and Cybernetics,\\
         10 ulyanova str.,Nizhny Novgorod 603005, Russia}\\
         e-mail:mel@cvsi.uucp.free.msk.su\\
         e-mail: kirillov@focus.nnov.su\\

\vskip 10mm
\end{center}
{\bf ABSTRACT}
\bigskip

\noindent
The problem of construction of a general ihomogeneous solution of
$D$-dimensional Einstein equations with any matter sources satisfying the
inequality $\epsilon \geq p$ in the vicinity of a cosmological
singularity is considered. It is shown that near the singularity a local
behavior of metric functions is described by a billiard on a space of
a constant negative curvature. The billiard is shown to have a finite volume
and consequently to be a mixing one. Dynamics of inhomogeneities of metric is
studied and it is shown that its statistical properties admit a complete
description. An invariant measure describing statistics of inhomogeneities is
obtained and a role of a minimally-coupled scalar field in dynamics of the
inhomogeneities is also considered.
\vskip 25mm

\centerline{Moscow 1994}
\pagebreak

\setcounter{page}{1}

\renewcommand{\thesection}{\arabic{section}}
\renewcommand{\theequation}{\arabic{section}.\arabic{equation}}
\catcode`\@=11 \@addtoreset{equation}{section}\catcode`\@=12

\section{Introduction}
As is well known a number of unified theories predict that dimension of the
Universe exceeds that of we normally experience at macroscopic level [1].
It is assumed that presently additional dimensions are hidden, for they are
compactified to the Planckian size, and they do not display themselves in
macroscopic
and even in microscopic processes. However, the situation must be changed as we
come back with time to the very beginning of the evolution of our Universe.
Standard cosmological models predict the existence of a singular point
at the very beginning and, therefore, the universe size could approach to the
Planckian scale.  Thus, in the early universe the additional dimensions, if
exist, must not be different from ordinary dimensions and should be taken
into account.  Moreover, one could expect that the existence of additional
dimensions may drastically change properties of the singularity and even
remove it.  The main aim of this paper is to construct a general solution of
multidimensional Einstein equations near a singularity and to investigate
properties of inhomogeneities.

The way to construct a general solution with singularity was indicated first
by Belinsky et al.in Ref. [2] for $D=4$,  where $D$
is the dimension of a spacetime.
Dynamics of metric at a particular point of space
was shown to resemble the behaviour of the well studied "mixmaster" (or  of
the
type-IX) homogeneous model and the last one has a complex stochastic nature
[3,4].
Subsequent utilizing of that construction has been done in Ref.[5] where
the so-called scalar-vector-tensor theory (or the case $D=5$) was considered
and
the main feature of the mixmaster model, i.e. the complex oscillatory
regime was shown to be also present in the 5-dimensional case.

An investigation of inhomogeneities of metric based on the general solutions
has
been considered first in Ref. [6]. The case of the scalar-tensor theory (or
$D=4+$ scalar fields) was considered and it turned out that the oscillatory
regime leads to the fractioning of the coordinate scale $\lambda$ of the
inhomogeneities of Kasner exponents ($\lambda \approx \lambda_0 2^{-N}$,
where $N$ is the number of elapsed Kasner epochs and $\lambda_0$ is the
initial scale of inhomogeneities).  However, the methods by means of which
the properties and statistics of the inhomogeneities were investigated turned
out to be unapplicable for general case (i.e. for the absence of scalar
fields as well as for the expanding universe). This problem has been solved
recently in Ref. [7].  In this paper we generalize the results obtained in
Ref. [7] to the  case of arbitrary number of dimensions $D$.

As it was mentioned above the main features of the dynamics of an inhomogeneous
gravitational field nearby the singularity in 4-dimensional case may be
summarized
as follows:

1. Locally dynamics of metric functions resembles the behaviour of the most
general
homogeneous "mixmaster" model [2], which has stochastic behaviour [3,4].
Just the stochastic behaviour leads to a monotonic decrease of the coordinate
scale of the metric inhomogeneities [6,7].

2. In the vicinity of a singularity a scalar field is the only kind  of matter
effecting the dynamics of metric [5].

These facts may be simply understood under the following qualitative estimates
(that is confirmed by subsequent consideration).
As is well known in cosmology the horizon size $l_h$ is a natural scale
measuring a distance from the singularity. Therefore, inhomogeneities may be
divided
into the large-scale ($l_i\gg l_h$) and small-scale ($l_i\ll l_h$) ones. The
horison size varies with time as $l_h\sim t$ (where $t$ is the time in
synchronous reference system) whereas the characteristic spatial dimension
of the inhomogeneity may be estimated as $l_i\sim t^{\alpha}$ (as $t\rightarrow
0$).
In a linear theory for an isotropic background the exponent $\alpha$ may be
expressed via the state equation of matter as $\alpha ={2\epsilon \over
3(p+\epsilon )}$ and what is important  $\alpha <1$. Thus, it is clear that
an arbitrary inhomogeneous field becomes large-scale in the sufficient
closeness to the singularity. Since the inhomogeneities are large-scale there
are no effects connected with propagating of gravitational waves etc, and
this would mean that inhomogeneities become passive.  Consequently, dynamics
of the field may be approximately described by the most general homogeneous
model depending parametrically upon the spatial coordinates. Note, however,
that the homogeneous model would appear to be in a general non-diagonal form.

The second fact may be understood in the same way.  As it was shown in Ref. [8]
the gravitational part of the Einstein equations at the singular point varies
with time, in the leading order, as $R^{\alpha}_{\beta}\sim t^{-2}$ whereas the
matter has the order $T^{\alpha}_{\beta}\sim t^{-2k}$, where $k$ depends upon
the state
equation as $k={\epsilon +p\over 2\epsilon }$. Thus, one can see that for the
equation of state satisfying the inequality $p<\epsilon $ we have $k<1$ and
only for the limited case $p=\epsilon$ the both sides turn out to be of the
same order.  We note that in the vicinity of a singularity scalar  fields
give just this equation of state.

As it is well known (see for example Ref.[5,9]) additional dimensions may be
treated
in ordinary gravity as a set of nonminimally coupled scalar and vector fields.
Therefore, one could expect that the main contribution in dynamics in the
vicinity
of a singularity would be given by those dynamical functions which are
connected with
scalar fields whereas other functions would play a passive role.

Thus, one could expect that in multidimensional cosmology local behaviour
of the metric functions (at a particular point of space) will be described by
a most general homogeneous model. Here, it is necessary to recall the important
property of the mixmaster universe that is the stochastic behaviour. The
problem of
stochasticity of homogeneous multidimensional cosmological models has been
investigated in a number of papers [10]. In particular, the
conclusion was made that chaos is absent in spaces whose dimension is
large enough. This negative result is, apparently, connected with the fact
that in the vicinity of a singularity the homogeneous model would appear to
be in a general nondiagonal form, instead of simple diagonal models
considered in Refs.[10]. Besides, recently in Ref.[11]
it was shown that a wide class of multidimensional models has stochastic
behaviour.  In this paper we shall show that stochastic behaviour is the
general property of multidimensional cosmology.

\section{Generalized Kasner Solution, Generalized Kasner Variables}

We consider the theory in canonical formulation. Basic variables are the
Riemann metric components  $g_{\alpha  \beta}$  with signature
($+,-,...,-$) and a scalar field $\phi$ specified  on  the n-manifold $S$,
and its conjugate momentum $\Pi^{\alpha \beta}=\sqrt{g} (K^{\alpha \beta}-
g^{\alpha \beta}K)$ and $\Pi_{\phi}$, where $\alpha =1,...,n$ and $K^{\alpha
\beta}$ is the extrinsic curvature of $S$. For the sake of simplicity we
shall consider $S$ to be compact i.e.  $\partial S=0$.  The action has in
Planck units the following form
\begin{equation} I=\int_{S}(\Pi^{ij}{\partial g_{ij}\over\partial
t}+\Pi_{\phi}
{\partial \phi \over \partial t} - NH^0-N_{\alpha
}H^{\alpha })d^{n}x dt, \end{equation} where
\begin{equation}
H^0={1\over\sqrt{g}}\left\{\matrix{ \Pi^{\alpha}_{\beta }          
\Pi_{\alpha}^{\beta }-{1\over
n-1}(\Pi^{\alpha}_{\alpha})^2 + {1\over 2}\Pi_{\phi}^2 +g(W(\phi )-R)}\right\},
\end{equation}
\begin{equation}
H^{\alpha}=-2\Pi^{\alpha\beta}_{|\beta}+g^{\alpha \beta }           
\partial _{\beta }\phi  \Pi _{\phi },
\end{equation}
here
\begin{equation}
W(\phi )={1\over 2}\left\{\matrix{ g^{\alpha \beta }                
\partial _{\alpha }\phi
\partial _{\beta }\phi  + V(\phi )}\right\}.
\end{equation}
A generalized Kasner solution is realized under the following assumption
\begin{equation}
\sqrt{g}T\sim (\Pi^{\alpha}_{\beta}, \Pi_{\phi}) \gg V=g(W-R),   
\end{equation}
where $\sqrt{g}T$ denotes the first three terms in (2.2).
Then, using (2.1) one can find the following solution of  the  multidimensional
Einstein equations
\begin{equation}
ds^{2}=dt^2 - \sum_{a=0}^{n-1} t^{s_a }l^a _{\alpha} \, ,          
l^a _{\beta}dx^{\alpha}dx^{\beta}
\end{equation}
where $l^a_{\alpha}$, $s_a$ are functions  of space coordinates. Kasner
exponents $s_a$ satisfy the identities $\sum s_a =\sum s_a ^2 +q^2 =1$,
and run the domain $-{n-2\over n }\leq s_a \leq 1$
(here $q^2 ={(n-1)^2\over 2}{\Pi _{\phi}^2\over (\Pi^{\alpha}_{\alpha})^2}$).
Since, as it was shown
in Ref.[2,5] the generalized Kasner solution takes a substantial portion of
the evolution of metric it is convenient to introduce a Kasner-like
parametrization of the dynamical variables [7].  We consider the following
representation for metric components and their conjugate momenta
\begin{equation}                     
g_{\alpha \beta }= \sum_{a} \exp \left\{ q^a \right\} l^a _{\alpha }l^a
_{\beta } \,  ,
\end{equation}
\begin{equation}                        
\Pi ^{\alpha}_{\beta }= \sum _{a} p_a L_a ^{\alpha }l^a _{\beta }\, ,
\end{equation}
here  $L_a ^{\alpha}l^b _{\alpha}=\delta^b _a $ ($a,b=0,...,(n-1)$),
and the vectors $l^a _{\alpha}$ contain only $n(n-1)$ arbitrary functions of
spatial coordinates. Further parametrization may be  taken  in  the  following
form
\begin{equation}                                                 
l^a _{\alpha}=U^a _b S^b _{\alpha}, \,U^a_b\in SO(n), \,S^a _{\alpha}=\delta^a
_{\alpha}+R^a _{\alpha}
\end{equation}
where $R^a _{\alpha}$ denotes a triangle matrix ($R^a _{\alpha}=0$ as $a
\leq\alpha$).
Substituting (2.7) - (2.9) into
(2.1) one gets the following expression for the action functional
\begin{equation}                                                
I=\int_{S}(p_a {\partial q^a\over\partial t}+T^{\alpha}_a {\partial
R_{\alpha}^a \over \partial t}
+\Pi_{\phi}{\partial \phi \over \partial t} - NH^0-N_{\alpha }H^{\alpha
})d^{n}x dt,
\end{equation}
here $T^{\alpha}_a =2\sum_b p_b L^{\alpha}_b U^b_a$ and the Hamiltonian
constraint
takes the form
\begin{equation}                                                
H^0={1\over\sqrt{g}}\left\{\matrix{ \sum p_a^2 -{1\over n-1}(\sum p_a)^2
+ {1\over 2}\Pi_{\phi}^2 +V }\right\}.
\end{equation}
In the case of $n=3$ the functions $R_{\alpha}^a $ are connected purely with
transformations of a coordinate system and may be removed  by  solving
momentum constraints $H^{\alpha}=0$. In  the multidimensional case the
functions $R_{\alpha}^a$ contain ${n(n-3)\over 2}$ dynamical functions as well.
Now it is easy to see that the choice of Kasner-like parametrization simplifies
the procedure of the constructing of the generalized Kasner solution. Indeed,
if we now
neglect the potential term in (2.10) and put $N^{\alpha}=0$ we find that
Hamiltonian
does not depend on the scale functions and other dynamical variables
contained in Kasner vectors introduced by expressions (2.7) (2.8).

\section{The asymptotic model in the vicinity of a cosmological singularity}

As it is well known, [2], [5], the Kasner regime (2.6) turns out to be unstable
in
a general case. This happens due to the violation of the condition (2.5)
because the
potential $V$ contains increasing terms which lead to replacement of Kasner
regimes.
To find out the law of replacement it is more convenient to use an asymptotic
expression
for the potential [7], [11]. For this aim we put the potential in the
following form \begin{equation}                           
V=\sum_{A=1}^{k} \lambda_{A} g^{u_{A}},
\end{equation}
here $\lambda _A$ is a set of functions of all dynamical variables and of their
derivatives and $u_a$ are linear functions of the anisotropy parameters
$Q_a ={q^a\over \sum q}$ ($u_A=u_A (Q)$).  Assuming  the finiteness of the
functions
$\lambda$ and considering the limit $g\rightarrow 0$ we find that the potential
$V$ may be modeled by potential walls
\begin{equation}                            
g^{u_A} \rightarrow \theta _{\infty }[u_A (Q)]=
\left\{ \begin{array}{ll} +\infty , \, u_A <0,
\\ 0 , \qquad u_a >0 \end{array} \right.
\end{equation}
Thus, putting $N^{\alpha}=0$ we can remove the passive dynamical function
$T^{\alpha}_a$, $R_{\alpha}^a$ from the action (2.10) and get
the reduced dynamical system
\begin{equation}                             
I=\int_{S}\left\{\matrix{p_a {\partial q^a\over\partial t}
+\Pi_{\phi}{\partial \phi \over \partial t} - \lambda \left\{ \sum p^2 -{1\over
n-1}(\sum p)^2 +{1\over 2}\Pi_{\phi}^2 +U(Q)\right\}
}\right\}d^{n}x dt,
\end{equation}
here $\lambda$ is expressed via the lapse function as
$\lambda={N\over\sqrt{g}}$.
In harmonic variables the action (3.3) takes the form formally coincided with
the action for a relativistic particle
\begin{equation}                             
I=\int_{S} \left\{\matrix{P_r {\partial z^r\over\partial t} - \lambda
^{'} (P_i ^2 +U -P_0 ^2) } \right\} d^{n}x dt, \end{equation} here
$r=0,...,n$, $i=1,...,n$, $q^a =A^a _j z^j +z^0$ ($j=1,...,n-1$), $z^n =
\sqrt{{n(n-1)\over 2}}\phi $ and the constant matrix $A^a _j$ obeys the
following conditions
\begin{equation}                             
\sum_{a} A^a _j =0, \, \sum_a A^a_j A^a_k =n(n-1)\delta_{jk}
\end{equation}
and can be expressed in the following form
$$
A^a _j =\sqrt{{n(n-1)\over j(j-1)}}(\theta^a _j -j\delta^a _j ),
$$
where $\theta ^a _j =
\left\{\begin{array}{ll}1
,  j>a\\
0, j\leq a \end{array} \right.$.

Since the timelike variable $z^0$ varies during the evolution as $z^0\sim \ln
g$
the positions of potential walls turn out to be moving. It is more convenient
to fix
the positions of walls. This may be done by using the so-called  Misner-Chitre
like
variables [11] ($\vec y =y^j$)
\begin{equation}                             
z^{0}=-e^{-\tau }{1+y^{2}\over
1-y^{2}}, \vec{z}=-2e^{-\tau }{\vec{y}\over 1-y^{2}},\qquad
y = \mid \vec{y} \mid <1.
\end{equation}
Using these variables one can find the following expressions for the anisotropy
parameters
\begin{equation}                             
Q_a (y)= {1\over n}\left\{\matrix{1+{2A^a _j y^j\over 1+y^2}}\right\},
\end{equation}
which are now independent of timelike variable $\tau$. From (3.7) one can find
the range of the anisotropy functions $-{n-2\over n}\leq Q_a\leq 1$.

Choosing as a time variable the quantity $\tau $ (i.e. in the gauge
$N={n(n-1)\over 2}\sqrt{g}\exp(-2\tau )/P^0 $) we put the action
(3.4) into the ADM form
\begin{equation}                              
I=\int_{S}
\left\{\matrix{\vec P {\partial \over \partial \tau }\vec y + P^n {\partial
\over \partial \tau } z^n
-P^0 (P,y)}\right\}
d^{n}xd\tau ,
\end{equation}
where the quantity
\begin{equation}                                  
P^0 (P,y)=(\epsilon ^{2}(\vec y,\vec P) + V[y] + (P^n) ^2 e^{-2\tau })^{1/2},
\end{equation}
plays the role of the ADM Hamiltonian density and
\begin{equation}                                  
\epsilon ^{2}={1\over 4}(1-y^{2})^{2}\vec P ^{2}.
\end{equation}

The part of the configuration space connected with the variables $\vec y$ is
a realization of the $(n-1)$ -dimensional Lobachevsky space [12] and the
potential $V$ cuts a part of it. Thus, locally (at a particular point of $S$)
the action (3.9) describes a billiard on the Lobachevsky space.  The
positions of walls which form the boundary of the billiard are determined,
due to (3.1) by the inequalities
\begin{equation}                             
\sigma _{abc}=1+Q_a -Q_b -Q_c \geq 0, \, a\ne b\ne c
\end{equation}
and the total number of walls is ${n(n-1)(n-2)\over 2}$.
Using the matrix (3.5) one can find that the walls are formed by
spheres determined by the equations
\begin{equation} 
\sigma_{abc}={n-1\over n(1+y^2 )}
\left\{\matrix{(\vec y+\vec B_{abc})^2 +1-B_{abc}^2}\right\}, \,
\vec B_{abc}={1\over n-1}(\vec A^a -\vec A^b -\vec A^c ),
\end{equation}
here for arbitrary $a,b,c$ we have $B^2 =1+{2n\over n-1}$.
In a general case $n$ points of the billiard having the coordinates
$\vec P_a={1\over n-1}\vec A^a$ lie on the absolute (at infinity of the
Lobachevsky space). Nevetheless, one can show that the volume of the
billiard is finite.  We give two simplest examples for illustration of the
billiards on fig.1. The case $n=3$ on fig.1a coincides with the well-known
"mixmaster" model and on fig.1b we illustrate the case of $n=4$ considered in
Ref.[5].

\section{Dynamics of inhomogeneities}
The system (3.8) has the form of the direct product of  "homogeneous"
local systems.
Each local system in (3.8) has two variables $\epsilon $ and $P^n $
as integrals of motion.  The solution of this local system for remaining
functions represents a geodesic flow on a manifold with negative curvature.
As it is well known the geodesic flow on a manifold with negative curvature
is characterized by exponential instability [12]. This means that during the
motion along a geodesic the normal deviations grow no slower than the
exponential of the traversed path $\xi \simeq \xi_{0}e^{s}$), where the
traversed path is determined by the expression
\begin{equation}
s=\int^{\tau }_{\tau _{0}}dl=\int^{\tau }_{\tau  _{0}}{2\mid {\partial
y\over \partial \tau} \mid \over (1- y ^2 )} d\tau = {1\over 2} \ln\mid {P^0
- \epsilon \over P^0 + \epsilon }\mid _{\tau  _{0}}^{\tau}.  \end{equation}
This instability leads to the stochastic nature of the corresponding geodesic
flow. The system possesses the mixing property [13]
and an invariant measure induced by the
Liouviulle one
\begin{equation}                                    
d\mu (y,P)= const \delta (E-\epsilon )d^{n-1}yd^{n-1}P,
\end{equation}
where $E$ is a constant. Integrating this expression over $\epsilon$  we find
\begin{equation}                                      
d\mu (y,s)= const {d^{n-1}yd^{n-2}s \over (1-y^{2})^{n}},
\end{equation}
where $\vec s={\vec P\over \epsilon}$, $|s|=1$.

Since the inhomogeneous system (3.8) is the direct product of "homogeneous"
systems one can simply describe its behaviour as in ref [7].
In particular, the scale of the inhomogeneity decreases as
\begin{equation}                
\lambda _i\sim ({\partial y\over\partial x})^{-1}\sim \lambda^0_i\exp
(-s) \end{equation} and after sufficiently large time ($s(\tau )\rightarrow
\infty$) the dynamical functions $\vec y(x)$, $\vec P(x)$ become a random
functions of the spatial coordinates.  Their statistics is described by the
invariant distribution (4.3) and asymptotic expressions for averages and
correlating functions have the form
\begin{equation}                
<\vec {y(x)}> =
<\vec {P(x)}> =0, \, <y_k (x),y_l (x')> = <y_k ,y_l >\delta (x,x'),
\end{equation} for $|x-x'|\gg\lambda^0_i\exp (-s)$.

Here it is necessary to point out a role of the scalar field in dynamics
and statistical properties of inhomogeneities. As may be easily seen from (4.1)
in the absence of a scalar field (i.e. $P^n=0$) the transversed path coincides
with the duration of motion (we have $s= \Delta \tau =\tau -\tau_0 $ instead of
(4.1)). Thus, the effect of scalar fields is displayed in the replacement of
the dependence for transversed path of time variable and, therefore, in the
replacement of the rate of increasing of the inhomogeneities. This
replacement does not change qualitatively the evolution of the universe in
the case of cosmological expansion.  But in the case of the contracting
universe the situation changes drastically.  Indeed, in the limit $\tau
\rightarrow -\infty$ from (4.1) we find that the transversed path $s$ takes a
limited value $s_0$ and therefore the increasing of inhomogeneities turns out
to be finite.  One of consequences of such behaviour is the fact that at the
singularity the functions $\vec y$ and $\vec P$ take constant values. In
other words in the presence of scalar fields a cosmological collapse ends
with a stable Kasner-like regime (2.6).  This fact may be seen in the other
way. Indeed, in the limit $\tau \rightarrow -\infty$ the scalar field gives
the leading contribution in ADM Hamiltonian  (3.9)
and  $P^0$  does  not  depend  on
gravitational variables at all.

The finiteness of the transversed path $s(\tau)$ leads, generally speaking, to
the destruction of the mixing  properties [13],  since  for  establishment
of the invariant measure it is necessary to satisfy the  condition
$s_{0}\rightarrow \infty$. Evidently, this condition requires the smallness
of the  energy density for scalar field as compared with the ADM energy of
gravitational field (the last term in (3.9) in comparison with the first
ones).  Indeed, in this case $s_{0}$ is determined by the expression $s_{0}=
- \ln{P^n  e ^{\tau _{0}}\over 2\epsilon }$ , which follows from (4.1), and
as $P^n \rightarrow 0$ one get $s_{0}\rightarrow \infty$ (i.e.  $s$ can have
arbitrary large values).

Thus, in the case of cosmological contraction one may speak of the mixing
and, therefore, of establishment of the invariant statistical distribution
just only for those spatial domains which have suficiently small energy
density  of  the scalar field.

\section{Estimates and concluding remarks}

In this manner the large-scale structure of the space in  the  vicinity  of
singularity   acquires a quasi-isotropic nature. A distribution    of
inhomogeneities is determined by the set of functions of spatial coordinates
$\epsilon(x)$, $\Pi_{\phi}(x)$  and $R^a _{\alpha}$ which conserve during the
evolution
a primordial degree of inhomogeneity of the space. The scale of inhomogeneity
of other functions grows as $\lambda \approx \lambda _{0}e^{-s(\tau )}$.
In this section we  give  some  estimates  clarifying  the  behaviour  of  the
inhomogeneities.  For simplicity we consider the case
when the scalar field is  absent.

To find the estimate for growth of the inhomogeneity in a synchronous time
$t$ $(dt=Nd\tau )$ we put $y=0$. Then for variation  of  the  variable  $\tau
$  one  may  find  the following estimate $\sqrt{g} \sim \exp(-{n\over
2}e^{-\tau })\sim P^0 t$,
(here the point $t=0$ corresponds to the singularity).
According to (4.4) the dependence of
the coordinate scale of inhomogeneity upon the time $t$ takes the form
$$
\lambda  \approx  \lambda _{0}\ln(1/g_{0})/\ln(1/g)
$$
in the case of contracting ($g\rightarrow 0$) and
$$
\lambda  \approx  \lambda _{0}\ln(1/g)/\ln(1/g_{0})
$$
in the case of the expanding universe.

A rapid generation of the more and more small scales leads to the formation of
spatial chaos in metric functions and so the large-scale structure acquires a
quasi-isotropic nature. Speeds of  the  scale  growing  (Hubble  constants) for
different directions turn out to be  equal  after  averaging  over  a  spatial
domains having the size $\approx \lambda_{0}$.
Indeed, using (3.7) one may find the expressions for averages $<Q_a>=1/n$.

Besides, it is necessary  to  mention
one more characteristic feature of the  oscillatory  regime  in the
inhomogeneous
case. This is the formation of a cellular structure in the scale functions
$Q_a$ during the
evolution which demonstrate explicitely the stochastic process of development
of inhomogeneities.  Indeed, let us consider some region of coordinate space
$\Delta V$.  Two functions $\vec {y(x)}$ define the map of that region on
some square $\Sigma \in K$ (see fig.1c).  During the evolution  the  size of
the square $\Sigma$ grows $\simeq e^{s(\tau )}$ and $\Sigma$ covers  the
domain of  the billiard $K$ many times. Each covering determines its own
preimage in $\Delta V$.  In this manner the initial coordinate volume is
splitted up in "cells" $\Delta V=\bigcup_{i}\Delta V_{i}$.  In the every cell
the vector $\vec{y(x)}$ takes almost  all admissible values $\vec y\in K$ and
that of the functions $Q_a$ (for $n=3$ $Q_a\in [0,1]$).  Such a structure
turns out to be depending of time and the number of  cells increases as
$N\approx N_{0}e^{s(\tau )}$.  However, the situation will be changed if we
consider a contracting space filled  with  a scalar field. Then the evolution
of this structure in the limit $g\rightarrow 0$ ends, because  the functions
$Q_a$ become independent of time, and on the final stage of the collapse one
would have a real cellular structure  [6].

In spite of the isotropic nature of the spatial distribution of the field
the large local anisotropy displays itself  in  the  anomalous  dependence  of
spatial lengths upon time variable for vectors and curves .
Indeed, in the case of $D=3$ a moment of scale function $<g^{MQ_a}>$ (where
$M>0$) is decreased in the asymptotic $g\rightarrow o$ as the Laplace
integral $\int^{1}_{0}g^{MQ_a}\rho (Q_a )dQ_a$, where $\rho (Q_a )$ is a
distribution which follows from (4.3) and has the form
\begin{equation}                         
\rho (H) = {2\over \pi } (Q(1 - Q))^{-1/2}(1 + 3Q)^{-1}.  \end{equation}
As $Q\ll 1$ one has $\rho (Q_a )\approx {2\over \pi }(Q_a )^{-1/2}$
and, thus, in the limit $g\rightarrow 0$ we get the estimate
\begin{equation}                          
<g^{MQ_a}> \approx  (M \ln(1/g))^{-1/2}.
\end{equation}
When $D>3$ the similar analysis may be done.

In conclusion we briefly repeat the main results.
The general ihomogeneous solution of
$D$-dimensional Einstein equations
with any matter sources satisfying the
inequality $\epsilon \geq p$
near the cosmological singularity is
constructed.  It is shown that near the singularity a local behavior of
metric functions ( at a particular point of the coordinate space) is
described by a billiard on the ($D-1$)-dimensional Lobachevsky space.  In
contrast to diagonal homogeneous models [10] in the inhomogeneous case the
billiard is shown to have always a finite volume and consequently to be a
mixing one. The rate of growth of inhomogeneities of metric is obtained.
Statistical properties of inhomogeneities are described by the invariant
measure.  It is shown that a minimally-coupled scalar field leads, in
general, to the distruction of stochastic properties of the inhomogeneous
model.

\begin{center}
Acknowledgments
\end{center}

This work was supported in part by the Russian Ministry of Science.

\end{document}